\documentclass[12pt,twoside]{article}
\usepackage{fleqn,espcrc1}

\usepackage{graphicx}

\newcommand{\AmS}{{\protect\the\textfont2
  A\kern-.1667em\lower.5ex\hbox{M}\kern-.125emS}}

\title{Precision determination of the $\pi $N scattering lengths and the
charged $\pi $NN coupling constant}

\author{T.E.O.~Ericson\address{TSL, Uppsala and CERN, Geneva}%
\thanks{Also parttime visitor at CSSM, Adelaide, S. Australia}
 B.~Loiseau\address{ LPNHE/LPTPE,
Universit\'{e}s D. Diderot and
P. \& M. Curie, Paris
and
TSL, Uppsala}%
and
 A.W.~Thomas\address{CSSM, University of Adelaide, Adelaide, S.
Australia}}

\begin{document}
 \maketitle

\begin{abstract} We critically evaluate the isovector GMO sumrule for the
charged $\pi N\! N$ coupling constant using recent precision data from
$\pi ^-$p and $\pi^-$d atoms and with careful attention to systematic
errors. From the $\pi ^-$d scattering length we deduce the pion-proton
scattering lengths $\frac {1}{2}(a_{\pi ^-p}+a_{\pi ^-n})=(-20\pm
6$(statistic)$ \pm 10$ (systematic))~$\cdot 10^{-4}m_{\pi _c}^{-1}$ and
$\frac
{1}{2}(a_{\pi ^-p}-a_{\pi ^-n})=(903 \pm 14)\cdot 10^{-4}m_{\pi
_c}^{-1}$.
From this a direct evaluation gives $g^2_c(GMO)/4\pi =14.20\pm
0.07$(statistic)$\pm 0.13$(systematic) or $f^2_c/4\pi = 0.0786\pm
0.0008$.
\end{abstract}

\section{INTRODUCTION.} The most direct and transparent way of determining
the $\pi $NN coupling constant directly from data is to use the basic
forward dispersion relation called the Goldberger-Miyazawa-Oehme (GMO)
sum-rule: 

\begin{equation}
 g_c^{2}/4\pi = -4.50 \cdot J^- +103.3 \cdot (\frac {a_{\pi ^- p}-a_{\pi
^+
p}}{2}), \end{equation}
 where $J^-$ is a weighted integral over the difference of $\pi ^{\pm}$p
total cross sections. The assumption of isospin symmetry is not necessary. 
The scattering lengths $a_{\pi ^{\pm} p}$ have become experimentally
accessible recently due to high precision experiments on $\pi ^-$p and
$\pi ^-$d atoms as will be discussed below. The main obstacle in the
evaluation of this sum rule in the past has been the necessity to rely, at
least in part, on scattering lengths constructed from higher energy $\pi
$N data via dispersion relations, a difficult procedure in view of the
high precision needed and the heavy cancellations in the symmetric
combination of the scattering lengths. We report here on a critical
precision evaluation of the scattering lengths using the deuteron
information and on the first direct evaluation of the GMO sum rule.

\section{DETERMINATION OF THE $\pi $N SCATTERING LENGTHS.} It is essential
to know the $\pi $N scattering lengths $a^+$
and
$a^-$ and
the $\pi $NN coupling coupling constant to high precision and with well
controlled uncertainties in order to make accurate tests of chiral
symmetry predictions. The potentially most accurate experimental source
for a determination of the isoscalar combination $a^+$ is the $\pi ^-$d
scattering length\cite{HAU98,BAD98}. To leading order this scattering
length is
given by the coherent sum

\begin {equation} a_{\pi ^-d}~=~a_{\pi ^-p}~+~a_{\pi ^-n}.
\end {equation}

 Since the two
components
cancel to nearly 1\% precision, even a rather summary description of the
deuteron gives a high absolute precision. The main correction is the
double scattering term, which in the static pointlike approximation is 
\begin {equation}D~=~2~\frac {(1+m_{\pi}/M)^2}{(1+m_{\pi}/M_d)}~[(\frac
{a_{\pi ^-p}+
a_{\pi^-n}}{2})^2-2(\frac {a_{\pi ^-p}-
a_{\pi^-n}}{2})^2]<1/r>,
\end {equation}
where $M_d$ is the deuteron mass. This term dominates heavily and its
approximate value is D~=~-0.0257 m$_{\pi }^{-1}$ to be compared to the
experimental $a_{\pi ^-d}~=~-0.0261~m_{\pi }^{-1}$. The most recent
theoretical investigation of the various correction terms inside multiple
scattering theory is due to Baru and Kudryatsev (B-K)\cite {BAR97a}; we
take their work as the departure for a critical and quantitative
assessment
of the theoretical uncertainties.  Typical results are given in Table 1. 

\begin{table}[htb]
\caption[h]{ Typical contributions to
a$_{\pi d}$ scattering length in units 10$^{-4}m_{\pi}^{-1}$.}
\begin{tabular}{|l|r|r|}
\hline
Contributions&Present work&B-K \cite {BAR97a}\\
\hline
a$_{\pi^- d}$(double scattering; static)  &-257(7) &-252\\
Fermi motion &61(7)&50\\
dispersion correction &-56(14) &not included\\
isospin violation &3.5&3.5\\
form factor  &23(6)&32(8)\\
higher orders  &4(1)&6\\
sp interference  &small &-44\\
non-static effects &11(6)&10\\
p wave double scattering   &-3 &-3\\
virtual pion scattering  &-7(2)&not considered\\
\hline
\hline
a$_{\pi d}$(experimental)&-261(5) &\\
\hline
\end{tabular}
\label{tab:contributions}
\end{table}

 The principal corrections are the following:

a) the Fermi motion of the nucleons, which produces a well defined
correction due to induced p wave scattering.

b) the dispersive correction due to the $\pi ^-d\rightarrow nn$
absorption.  This term has been evaluated in 3-body Fadeev
calculations\cite {MIZ77}.  It is the single largest source of
uncertainty.
This correction was not included in B-K.

c) B-K advocate a rather large correction from 'sp' interference, in which
one scattering is s wave and the other one p wave due to 'Galilean
invariant' off mass shell contributions. The procedure is model dependent. 
We find that the dominant contribution is generated by the spin averaged
isovector p wave $\pi N$ Born term, which has no ambiguity in its
of-mass-shell structure. The
contribution vanishes for this term and no correction should be made for
this effect. This is our most important change from the results of B-K.

d) corrections for non-pointlike interactions and form factors. No major
uncertainty is introduced by these effects with liberal variation of
parameters.

e) non-static effects.  These produce only rather small effects. 

f) corrections for virtual pion scattering, isospin violation, p wave
double scattering and higher order terms are all small and controllable 
corrections.

In particular, the assumption of
isospin symmetry is only needed in correction terms and, on the level of
expected violations, has a negligible influence on the conclusions. 

Based on this, we obtain well controlled values for the symmetric and
antisymmetric combinations of scattering lengths
 $\frac {1}{2}(a_{\pi ^-p}\pm a_{\pi ^-n})~\simeq ~ a^{\pm }$ deduced from
the
data. The preliminary values are

\begin{equation}\frac {a_{\pi ^-p}+a_{\pi
^-n}}{2}=(-0.0020 \pm 10\pm 6)m_{\pi}^{-1};~~\frac {a_{\pi ^-p}-
a_{\pi
^-n}}{2}=(0.0903 \pm 10\pm 9)m_{\pi}^{-1}.\end{equation}

Here the first error is systematic and the second one statistical.  As
seen from the Figure these values are in excellent agreement with the
value based on the width of the pionic hydrogen 1s state\cite{BAD98}, but
the uncertainty is presently much smaller.\\
\includegraphics[angle=90,width=10cm]{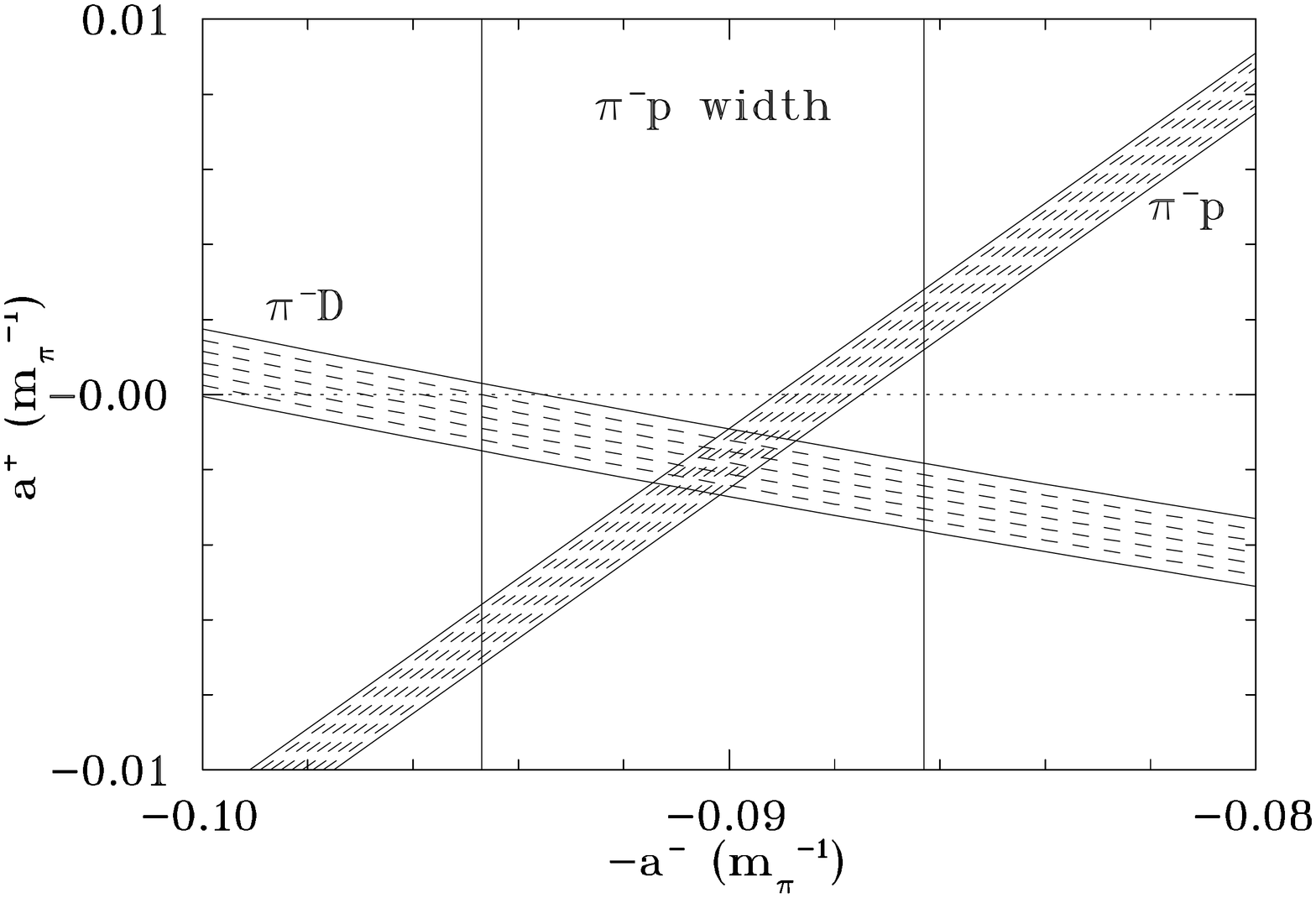}
\section{EVALUATION OF THE $\pi $NN COUPLING CONSTANT.}  We report as
well the first evaluation of the GMO sum rule
based
directly on observables.  This gives  a controlled and
model-independent value
for the the charged coupling constant $g_c^2$.  We rewrite the sum rule in
the following convenient and robust form, which emphasizes its directly
determined empirical ingredients: 

 \begin
{equation}
 g_c^{2}/4\pi =
-4.50 \cdot J^- +103.3 \cdot a_{\pi ^- p}-103.3 \cdot (\frac {a_{\pi ^-
p}+a_{\pi ^+ p}}{2}).
\end{equation}

Here the total cross section integral $J^-$ is in mb and the scattering
lengths in units of $m_{\pi _c}^{-1}$. The systematic error in J$^-$
entering this relation is presently a major limitation on its accuracy of
eq. (5) and we have critically examined its contributions. We find $J^-$
=-1.083(25)mb, which agrees well with values used in previous evaluations
of the sum rule, but with errors under control. 
 Assuming charge symmetry, i.e., $a_{\pi ^+p}=a_{\pi ^-n}$, we find from
eq.(4) above the following preliminary value for the charged coupling
constant

\begin {equation}
g^2_c/4\pi =4.87(11)+9.12(8)+0.21(12)=14.20(18).\end {equation}

Since this value is directly based on data it supersedes previous
evaluations of the sumrule.  Our value for $g^2_c/4\pi$ with systematic
errors
under control is only with difficulty consistent with the low values in
the range 13.5-13.6 advocated in ref.\cite{SWA97}, since it differs by
over 3 standard deviations. There is a possible discrepancy by 2 standard
deviations with the value 13.75(9) from $\pi $N dispersion
analysis\cite{Arndt:1998zd}. The origin is not clear. This group evaluates
the scattering lengths within the framework and find the same value 13.75
from their GMO relation. The heavy cancellation of the components in the
term $1/2(a_{\pi ^-p}+a_{\pi ^-n})$ is however quite subtle to reproduce
to high precision in this dispersive approach and the value they find
would represent a 50\% contribution to the deuteron scattering length.
This contradicts the more direct deuteron result of eq.  (4).  This might
indicate a too strong reliance on isospin symmetry in the dispersive
analysis, since this extremely small amplitude is less than 1:$10^3$ of
the amplitude in the resonant region. On the other hand, our result is
consistent to 1 standard deviation with the value from np charge exchange
14.52(26)\cite{RAH98}. It is also consistent to one standard deviation
with the Goldberger-Treiman relation which gives a coupling constant
13.99(17)  assuming that its discrepancy is due entirely to a $\pi $NN
monopole form factor with a cut-off of 800 MeV/c as typically found from
the Cloudy Bag Model\cite{GUI83}, meson-theoretical
ones\cite{Bockmann:1999nu} etc.. \\ \\

\end{document}